\documentclass{desyproc}

\begin{document}
\title{Axion BEC Dark Matter}

\author{{\slshape Ozgur Erken, Pierre Sikivie,
Heywood Tam, Qiaoli Yang}\\[1ex]
Department of Physics, University of Florida, 
Gainesville, FL 32611, USA}

\contribID{familyname\_firstname}

\desyproc{DESY-PROC-2011-04}
\acronym{Patras 2011} 
\doi  

\maketitle

\begin{abstract}
Cold dark matter axions thermalize through gravitational self-interactions 
and form a Bose-Einstein condensate when the photon temperature reaches
approximately 500 eV.  Axion Bose-Einstein condensation provides an 
opportunity to distinguish axions from the other dark matter candidates 
on the basis of observation.  The rethermalization of axions that are about 
to fall in a galactic potential well causes them to acquire net overall 
rotation, whereas ordinary cold dark matter falls in with an irrotational 
velocity field.  The inner caustics of galactic halos are different in the 
two cases.

\end{abstract}

\section{Introduction}

Both axions and weakly interacting particles (WIMPs) are considered forms of 
cold dark matter (CDM).  Furthermore, until recently, axions and WIMPs were 
thought to be indistinguishable on observational grounds, i.e. indistinguishable 
on the basis of purely astronomical data.  The discovery \cite{CABEC} that dark 
matter axions form a Bose-Einstein condensate (BEC) has changed this view since 
axion BEC is claimed to have observable consequences \cite{CABEC,case,Li7}.  
This raises the question:  When do axions behave as ordinary CDM  and when 
do they not?

To start off it is worth emphasizing that, at the fundamental level,
axions and WIMPs are very different.  The surprise is really that they
have similar properties as far as large scale structure is concerned.
Both axions and WIMPs are described by quantum fields.  Furthermore,
both are excellently described by classical limits of quantum fields.
But the classical limits are different in the two cases:  WIMPs are in
the classical particle limit whereas (decoupled) axions are in the classical
field limit.  In the classical particle limit one takes $\hbar \rightarrow 0$
while keeping $E = \hbar \omega$ and $\vec{p} = \hbar \vec{k}$ fixed.  Since
$\omega, \vec{k} \rightarrow \infty$, the wave nature of the quanta disappears.
WIMPs are to excellent approximation classical point particles.  In the
classical field limit, on the other hand, one takes $\hbar \rightarrow 0$
for constant $\omega$ and $\vec{k}$.  $E = {\cal N} \hbar \omega$ and
$\vec{p} = {\cal N} \hbar \vec{k}$ are held fixed by letting the quantum
state occupation number ${\cal N} \rightarrow \infty$.  This is the limit
in which quantum electrodynamics becomes classical electrodynamics.  It is
the appropriate limit for (decoupled) cold dark matter axions because they
are a highly degenerate Bose gas.  The axion states that are occupied have
huge occupation numbers, ${\cal N} \sim 10^{61}$ \cite{CABEC}.  The need
to restrict to {\it decoupled} axions will be explained shortly.

So axions and WIMPs are fundamentally different even if both can 
legitimately be called CDM.  The distinction is not just academic, 
and is certainly important if axions thermalize, i.e. if axions
find a state of larger entropy through self-interactions.  Recall that,
whereas statistical mechanics makes sense of the behaviour of large
aggregates of classical particles (it was invented by Boltzmann to derive
the properties of atoms in the gaseous state) it fails to make sense of
classical fields. In thermal equilibrium every mode of a classical
field would have average energy $k_{\rm B} T$.  As Rayleigh pointed out,
the energy density is infinite then at finite temperature due to the
contributions from short wavelength modes.  Thus the application of
statistical mechanics to classical field theory (classical electrodynamics
in particular) is in direct disagreement with observation.  As is well-known,
the disagreement is removed because of, and only because of, quantum mechanics.

In summary, if the axions are decoupled (i.e. do not interact and hence do not
thermalize), they behave to excellent approximation like a classical field.  
However, if the axions thermalize, they are not described by a classical field.  
Instead they form a Bose-Einstein condensate, an essentially quantum-mechanical 
phenomenon.

\section{Axion Bose-Einstein Condensation}

Axions were originally introduced \cite{PQWW} as a solution to the 
strong CP problem.  It was later found that they are a cold dark 
matter candidate \cite{axdm}.  Cold axions are produced when the 
axion mass turns on during the QCD phase transition.  The critical 
time, defined by $m(t_1) t_1 = 1$, is 
$t_1 \simeq 2 \cdot 10^{-7}~{\rm sec}~(f / 10^{12}~{\rm GeV})^{1 \over 3}$, 
where $f$ is the axion decay constant.  The zero temperature axion mass is 
given in terms of $f$ by
\begin{equation}
m \simeq 6 \cdot 10^{-6}~{\rm eV}~{10^{12}~{\rm GeV} \over f}~~\ .
\label{mass}
\end{equation}
At temperatures well above 1 GeV, the axion mass is practically zero.
It increases from zero to $m$ during the QCD phase transition.  The 
cold axions are the quanta of oscillation of the axion field that result 
from the turn on of the axion mass.  They have number density \cite{axdm}
\begin{equation}
n(t) \sim {4 \cdot 10^{47} \over {\rm cm}^3}~
\left({f \over 10^{12}~{\rm GeV}}\right)^{5 \over 3}
\left({a(t_1) \over a(t)}\right)^3
\label{numden}
\end{equation}
where $a(t)$ is the cosmological scale factor.  Because the axion
momenta are of order ${1 \over t_1}$ at time $t_1$ and vary with
time as $a(t)^{-1}$, the velocity dispersion of cold axions is
\begin{equation}
\delta v (t) \sim {1 \over m t_1}~{a(t_1) \over a(t)}
\label{veldis}
\end{equation}
{\it if} each axion remains in whatever state it is in, i.e. if axion
interactions are negligible.  We refer to this case as the limit of
decoupled cold axions.  If decoupled, the average state occupation 
number of cold axions is
\begin{equation}
{\cal N} \sim~ n~{(2 \pi)^3 \over {4 \pi \over 3} (m \delta v)^3}
\sim 10^{61}~\left({f \over 10^{12}~{\rm GeV}}\right)^{8 \over 3}~~\ .
\label{occnum}
\end{equation}
Clearly, the effective temperature of cold axions is much smaller than
the critical temperature
\begin{equation}
T_{\rm c} = \left({\pi^2 n \over \zeta(3)}\right)^{1 \over 3}
\simeq 300~{\rm GeV}~\left({f \over 10^{12}~{\rm GeV}}\right)^{5 \over 9}~
{a(t_1) \over a(t)}
\label{Tc}
\end{equation}
for BEC.  Axion number violating processes, such as their decay to two
photons, occur only on time scales vastly longer than the age of the
universe.  The only condition for axion BEC that is not manifestly
satisfied is thermal equilibrium.

Axions are in thermal equilibrium if their relaxation rate $\Gamma$ 
is large compared to the Hubble expansion rate $H(t) = {1 \over 2t}$.  
However, the usual techniques of non-equilibrium statistical mechanics 
are not applicable to dark matter axions.  On the one hand, cold axions 
are highly condensed in phase space (${\cal N} >> 1$), which greatly 
exaggerates the quantum effect of Bose-enhancement in their scattering 
processes. On the other, because their energy dispersion is very small, 
they are outside the `particle kinetic regime'. The picture of instantaneous 
collisions breaks down, and the usual Boltzmann equation no longer applies.  
The particle kinetic regime is defined by  $\delta \omega >> \Gamma$ where 
$\delta \omega$ is the energy dispersion of the particles.  Axions are in 
the opposite regime, $\delta \omega << \Gamma$, which we call the `condensed 
regime'.  Thermalization in the condensed regime is discussed in detail in 
ref. \cite{therm}, which gives an estimate for the relaxation rate of cold 
dark matter axions due to their gravitational self-interactions:
\begin{equation}
\Gamma \sim 4 \pi G n m^2 \ell^2
\label{rate}
\end{equation}
where $\ell \sim (m \delta v)^{-1}$ is the axion correlation length.
$\Gamma(t)/H(t)$ is of order
$5 \cdot 10^{-7}(f/10^{12}~{\rm GeV})^{2 \over 3}$
at time $t_1$ but grows as $t a^{-1}(t) \propto a(t)$.  Thus
gravitational interactions cause the axions to thermalize and
form a BEC when the photon temperature is of order
500 eV~$(f/10^{12}~{\rm GeV})^{1 \over 2}$.
 
The question is then whether axion BEC has observable consequences.
It is shown in refs. \cite{CABEC,therm} that cold dark matter axions 
behave as ordinary cold dark matter on all scales of observational 
interest when they are non-interacting.  Observable differences between 
cold axions and ordinary CDM occur only when the axions self-interact
or interact with other species.  Before Bose-Einstein condensation, cold 
axions are described by a free classical field and are indistinguishable 
from ordinary cold dark matter on all scales of observational interest.  
After Bose-Einstein condensation, almost all axions are in the same state.  
In the linear regime of evolution of density perturbations and within the 
horizon, the lowest energy state is time independent and no rethermalization 
is necessary for the axions to remain in the lowest energy state.  In that 
case, axion BEC and ordinary CDM are again indistinguishable on all scales 
of observational interest \cite{CABEC}.  However, beyond first order 
perturbation theory and/or upon entering the horizon, the axions 
rethermalize to try and remain in the lowest energy available 
state.  Axion BEC behaves differently from CDM then and the 
resulting differences are observable.

An example of an observable distinction between axion BEC and ordinary 
CDM is given in the next section.  Another possible observable distinction 
is the cooling of cosmic photons by thermal contact with the axion BEC.  If 
this happens, the baryon-to-photon ratio at nucleosynthesis and the effective 
number of neutrinos (a measure of the radiation density at recombination) are 
modified compared to their values in the standard cosmological model \cite{Li7}.

\section{Tidal torquing with axion BEC}

Let us consider axion BEC dark matter as it is about to fall into the 
gravitational potential well of a galaxy.  The gravitational field of 
neighbouring galaxies applies a tidal torque \cite{TTT} to the axion BEC.
Under what conditions is thermalization by gravitational self-interactions 
sufficiently fast that the condensed axions remain in the lowest energy 
available state as the space-time background evolves?  Following the 
arguments or ref. \cite{therm}, we expect that the axion BEC rethermalizes 
provided the gravitational forces produced by the BEC are larger than the 
typical rate $\dot{p}$ of change of axion momenta required for the axions 
to remain in the lowest energy state.  The gravitational forces are of order 
$4 \pi G n m^2 \ell$.  In this case, the correlation length $\ell$ must be 
taken to be of order the size $L$ of the region of interest since the 
gravitational fields due to axion BEC outside the region do not help the
thermalization of the axions within the region.  Hence the condition is
\begin{equation}
4 \pi G n m^2 L \gtrsim \dot{p}~~~\ .
\label{thercon}
\end{equation}
The self-similar infall model \cite{selfsim} was used \cite{therm}
to estimate $L$ and $\dot{p}$ as functions of time.  Furthermore, 
assuming that most of the dark matter is axions, the Friedmann 
equation implies
\begin{equation}
4 \pi G n m \simeq {3 \over 2} H(t)^2 \simeq {2 \over 3 t^2}
\label{Fried2}
\end{equation}
after equality between matter and radiation.  It is found \cite{therm} 
that Eq.~(\ref{thercon}) is satisfied at all times from equality till 
today by a margin of order 30.

We conclude that the axion BEC does rethermalize before falling
into the gravitational potential well of a galaxy.  Most axions
go to the lowest energy state consistent with the total angular
momentum acquired from neighboring inhomogeneities through tidal
torquing \cite{TTT}.  That state is a state of rigid rotation on
the turnaround sphere, implying $\vec{\nabla} \times \vec{v} \neq 0$
where $\vec{v}$ is the velocity field of the infalling axions.  In
contrast, the velocity field of WIMP dark matter is irrotational.
The inner caustics of galactic halos are different in the two cases.
Axions produce caustic rings \cite{crdm,sing} whereas WIMPs produce
the `tent-like' caustics described in ref.~\cite{inner}.  There is 
evidence for the existence of caustic rings in various galaxies at 
the radii predicted by the self-similar infall model.  For a review 
of this evidence see ref. \cite{MWhalo}.  It is shown in ref. \cite{case} 
that the phase space structure of galactic halos implied by the evidence
for caustic rings is precisely and in all respects that predicted by the 
assumption that the dark matter is a rethermalizing BEC.

\section{Acknowledgments}

This work was supported in part by the U.S. Department of Energy 
under grant DE-FG02-97ER41209.

\section{Bibliography}

\begin{footnotesize}

\end{footnotesize}


\begin{thebibliography}{99}

\bibitem{CABEC}
P. Sikivie and Q. Yang, Phys. Rev. Lett. 103 (2009) 111301.

\bibitem{case}
P. Sikivie, Phys. Lett. B 695 (2011) 22.

\bibitem{Li7} 
O. Erken, P. Sikivie, H. Tam and Q. Yang, arXiv:1104.4507.

\bibitem{PQWW}
R.D. Peccei and H. Quinn, Phys. Rev. Lett. 38 (1977) 1440 and
Phys. Rev. D16 (1977) 1791; S. Weinberg, Phys. Rev. Lett. 40 (1978) 
223; F. Wilczek, Phys. Rev. Lett. 40 (1978) 279.

\bibitem{axdm}
J. Preskill, M. Wise and F. Wilczek, Phys. Lett. B120 (1983) 127;
L. Abbott and P. Sikivie, Phys. Lett. B120 (1983) 133;
M. Dine and W. Fischler, Phys. Lett. B120 (1983) 137.

\bibitem{therm}
O. Erken, P. Sikivie, H. Tam and Q. Yang, arXiv:1111.1157.

\bibitem{TTT}
P.J.E. Peebles, Ap. J. 155 (1969) 2, and Astron. Ap. 11 (1971) 377.

\bibitem{selfsim}
J.A. Fillmore and P. Goldreich, Ap. J. 281 (1984) 1;
E. Bertschinger, Ap. J. Suppl. 58 (1985) 39;
P. Sikivie, I. Tkachev and Y. Wang, Phys. Rev. Lett. 75
(1995) 2911; Phys. Rev. D56 (1997) 1863.

\bibitem{crdm}      
P. Sikivie, Phys. Lett. B432 (1998) 139.

\bibitem{sing}
P. Sikivie, Phys. Rev. D60 (1999) 063501.

\bibitem{inner}
A. Natarajan and P. Sikivie, Phys. Rev. D73 (2006) 023510.

\bibitem{MWhalo}
L. Duffy and P. Sikivie, Phys. Rev. D78 (2008) 063508.

\end{thebibliography}
\end{document}